\documentclass[reprint,twocolumn,amsmath,amssymb,floatfix,nofootinbib,longbibliography,superscriptaddress]{revtex4-2}

\usepackage[colorlinks=true,citecolor=blue,linkcolor=magenta]{hyperref}
\usepackage{soul}
\usepackage{dsfont}
\usepackage{amsmath,amssymb}
\DeclareMathOperator{\Tr}{Tr}
\usepackage[normalem]{ulem}
\usepackage{comment}
\usepackage{graphicx}
\graphicspath{{plots/}}
\usepackage{braket}
\usepackage[usenames]{color}
\usepackage{float}
\usepackage{xcolor}
\newcommand{\NF}{\mathcal{N}_{F}}

\newcommand{\dt}{\Delta_t}

\newcommand{\di}{\Delta_I}
\newcommand{\Lm}{\Lambda}
\newcommand{\mt}[1]{\mathcal{#1}}
\newcommand{\Fv}{\mt{F}_{\text{vol}}}
\newcommand{\Fl}{\mt{F}_{\text{lin}}}
\newcommand{\pd}{\phantom\dagger}
\newcommand{\comm}[1]{}

\def \titlename {Scaling of Fock-space propagator and multifractality across the many-body localization transition
}

\begin{document}

\title{\titlename}
\author{Jagannath Sutradhar$^\clubsuit$}
\email{sjagannath@iisc.ac.in}
\affiliation{Centre for Condensed Matter Theory, Department of Physics, Indian Institute of Science, Bangalore 560012, India}

\author{Soumi Ghosh$^\clubsuit$}
\email{soumighosh@alum.iisc.ac.in}
\affiliation{Centre for Condensed Matter Theory, Department of Physics, Indian Institute of Science, Bangalore 560012, India}
\affiliation{International Centre for Theoretical Sciences, Tata Institute of Fundamental Research, Bengaluru 560089, India}
\author{Sthitadhi Roy}
\email{sthitadhi.roy@icts.res.in}
\affiliation{International Centre for Theoretical Sciences, Tata Institute of Fundamental Research, Bengaluru 560089, India}
\affiliation{Physical and Theoretical Chemistry, Oxford University, South Parks Road, Oxford OX1 3QZ, United Kingdom}
\affiliation{Rudolf Peierls Centre for Theoretical Physics, Clarendon Laboratory, Oxford University, Parks Road, Oxford OX1 3PU, United Kingdom}
\author{David E. Logan}
\email{david.logan@chem.ox.ac.uk}
\affiliation{Physical and Theoretical Chemistry, Oxford University, South Parks Road, Oxford OX1 3QZ, United Kingdom}
\affiliation{Department of Physics, Indian Institute of Science, Bangalore 560012, India}
\author{Subroto Mukerjee}
\email{smukerjee@iisc.ac.in}
\affiliation{Centre for Condensed Matter Theory, Department of Physics, Indian Institute of Science, Bangalore 560012, India}
\author{Sumilan Banerjee}
\email{sumilan@iisc.ac.in}
\affiliation{Centre for Condensed Matter Theory, Department of Physics, Indian Institute of Science, Bangalore 560012, India}
% 	\author{\authornames}
% 	\thanks{equally contributed}
% 	\affiliation{\affiliations}
% 	\email{sjagannath@iisc.ac.in}
% 	\email{soumighosh@alum.iisc.ac.in}
% 	\email{sthitadhi.roy@icts.res.in}
% 	\email{david.logan@chem.ox.ac.uk}
% 	\email{smukerjee@iisc.ac.in}
% 	\email{sumilan@iisc.ac.in}
% 	\date\today
	
\begin{abstract}
We implement a recursive Green function method to extract the  Fock space (FS) propagator and associated self-energy
across the many-body localization (MBL) transition, for one-dimensional interacting fermions in a random onsite potential. We show that the typical value of the imaginary part of the local FS self-energy, $\Delta_t$, related to the decay rate of an initially localized state, acts as a probabilistic order parameter for the thermal to MBL phase transition; and can be used to characterize critical properties of the transition as well as the multifractal nature of MBL states as a function of disorder strength $W$. In particular, we show that a fractal dimension $D_s$ extracted from $\dt$ jumps discontinuously across the transition, from $D_s<1$ in the MBL phase to $D_s= 1$ in the thermal phase. Moreover, $\dt$ follows an asymmetrical finite-size scaling form across the thermal-MBL transition, where a non-ergodic volume in the thermal phase diverges with a Kosterlitz-Thouless like essential singularity at the critical point $W_c$, and controls the continuous vanishing of $\dt$ as $W_{c}$ is approached. In contrast, a correlation length ($\xi$) extracted from $\dt$ exhibits a power-law divergence on approaching $W_c$ from the MBL phase. 
\end{abstract}

\maketitle
\def\thefootnote{$\clubsuit$}\footnotetext{These authors contributed equally to this work}\def\thefootnote{\arabic{footnote}}
%%%%%%=============================================
\section{Introduction}
A many-body localized (MBL) phase is a fascinating non-equilibrium state of matter which can originate in isolated quantum systems in the presence of disorder and interactions~\cite{Huse_review.2015,Abanin_review.2017,Alet_review.2018,RMP_MBLDmitry_2019}. In the MBL phase, high-energy eigenstates violate the Eigenstate Thermalization Hypothesis (ETH) \cite{Deutsch1991,Srednicki1994,Srednicki1999}, and local memory can be retained up to arbitrarily long times under time evolution. After the early landmark papers \cite{Basko2006,Gornyi2005},
a mathematical proof \cite{Imbrie2016} and numerous numerical studies \cite{Oganesyan.2007,Znidaric.2008,Pal.2010,Bardarson2012,Kjall.2014,Serbyn.2015} have provided strong evidence in favour of the existence of the MBL phase
in one dimension. However, the universal properties of the transition from the thermal or ETH phase to the MBL phase 
remain under active debate~\cite{Luitz.2015,Dumitrescu_KTMBL2019,Maximilian_PRL2020,suntajs2020quantum,panda2020can,abanin2021distinguishing,Piotr2020}. This is mainly due to the challenge to  ``first principles'' numerical verification of theoretical scenarios, posed by the exponentially growing  Fock-space dimension ($\mathcal{N}_F$) with system size $L$~ \cite{Oganesyan.2007,Pal.2010,Luitz.2015,Modak.2015,Khemani.2017}.  To complement calculations on microscopic models, several phenomenological renormalisation group-based approaches have been employed ~\cite{Goremykina_PRL2019,Dumitrescu_KTMBL2019,morningstar2019rg,Morningstar_PRB2020}, which  suggest a Kosterlitz-Thouless (KT)-like scenario for the MBL transition.

A further complementary approach to MBL is to consider it as an effective `Fock-space (FS) localization' problem 
of a fictitious particle on the complex, correlated FS graph (or `lattice')~\cite{altshuler1997quasi,Serbyn.2015,Logan.2019,roy2020fock}. This has led to crucial insights such as the role of strong FS correlations in stabilising the MBL phase~\cite{Altland2017,Logan.2019,Ghosh.2019,roy2020fock}, multifractality of eigenstates therein~\cite{deluca2013,Luitz.2015,Nicolas_MBL2019,roy.2021,DeTomasiRare2021} and, importantly, a numerical scaling theory of the MBL transition in terms of FS inverse participation ratios (IPR) of eigenstates, which is consistent with a KT-like scenario~\cite{Nicolas_MBL2019,roy.2021}. Here we ask the question: can a scaling theory of the MBL transition be formulated in terms of FS propagators? 

In this work we answer this question in the affirmative, by studying the propagator or Green function on the FS lattice for a fermionic chain with $L\leq 22$. In particular, we extract the local Feenberg self-energy \cite{Feenberg_1948,Logan.2019,Economoubook} from the diagonal elements of the FS propagator. As known since Anderson's seminal paper \cite{Anderson1958}, the typical value of the imaginary part of the Feenberg self-energy, $\dt$, acts as a probabilistic order parameter in the thermodynamic limit for Anderson transitions~\cite{Anderson1958,AbouChacra1973,economous1972existence,licciardello1975study}. This quantity has recently been employed to construct a self-consistent mean-field theory of the MBL transition on the Fock space~\cite{Logan.2019,roy2020fock}.
However, numerically exact evaluation of the FS self-energy for the microscopic models of MBL, and analysis of its behaviour across the MBL transition, is scarce. Here we fill this void and show that the MBL transition from ergodic extended states in the thermal phase to multifractal states in the MBL phase~\cite{deluca2013,Luitz.2015,Bertrand_2016,Nicolas_MBL2019,roy.2021,DeTomasiRare2021,Torres2017,Serbyn2017}, is manifest in an anomalous scaling of $\dt$ with the Fock-space dimension. This, 
together with a scaling theory of the MBL transition based on the order parameter $\Delta_t$, that is consistent with the KT-like scenario, constitutes the central result of the work.

The FS propagator contains information about the ergodic/non-ergodic nature of the phase, as different eigenstates contribute to it, based on their energy and amplitudes over FS lattice sites. From a technical point of view, computation of the propagator in principle requires all eigenstates. This restricts the system sizes accessible to numerical exact diagonlization (ED), which for $L>18$ can access only a limited number of eigenstates~\cite{Luitz.2018}. To access larger sizes, comparable to those accessible via parallelised shift-invert method~\cite{Luitz.2018} or POLFED~\cite{polfed}, but at significantly cheaper computational cost, we compute the FS propagators using a standard recursive Green function method \cite{Lee.1981,MacKinnon.1980,MacKinnonKramer.1983}, but adapted to the FS graph. Moreover, for the reasons discussed below, we study a scaled version of the self-energy, viz.\ $\dt/\sqrt{L}$, and refer to it as $\Delta_t$ throughout the rest of the paper for notational convenience. Employing the recursive method, we obtain the following main results:\\
1. The typical value $\Delta_t$ of the self-energy is finite in the thermal phase. It vanishes 
$\propto \mathcal{N}_F^{-(1-D_s)}$ in the MBL phase and at the critical point, where $D_s<1$ is a fractal dimension reflecting the multifractal nature of the states.  $D_{s}$ changes discontinuously across the MBL transition, from 
$D_s<1$ to $D_s=1$ throughout the thermal phase. \\
2. The finite-size scaling of $\dt$ as a function of disorder strength ($W$) is consistent with an asymmetric finite-size scaling form~\cite{Garc_PRL2017,Nicolas_MBL2019,Laflorencie2020,roy.2021} across the  MBL transition. 
Scaling on the thermal side is controlled by a non-ergodic volume scale $\Lambda$, which diverges with an essential singularity $\Lambda\sim \exp{(b/\sqrt{\delta W})}$,  [$\delta W=(W_c-W)$, $b\sim\mathcal{O}(1)$] at a critical disorder ($W_c$), redolent of a KT-like transition. Scaling on the MBL side by contrast is controlled by a FS correlation length ($\xi$), which exhibits a power-law divergence on approaching criticality. Moreover, the scaling theory implies that  in the thermodynamic limit $\dt$ vanishes continuously on approaching the transition from the thermal side as $\sim \exp{[-b'/\sqrt{\delta W}]}$ with $b'\sim \mathcal{O}(1)$.

As already mentioned, multifractal characterization of MBL states~\cite{deluca2013,Luitz.2015} and numerical scaling theory consistent with a KT-like MBL transition have been obtained via study of eigenstate 
IPRs~\cite{Nicolas_MBL2019,roy.2021}. However our work reveals for the first time the multifractality and KT-type critical scaling in terms of a FS order parameter for the thermal-MBL transition. Additionally, the FS order parameter, being associated with an inverse decay time of localized initial states, provides a truly dynamical characterization of MBL transition unlike the static properties studied based on eigenstates in the previous studies \cite{deluca2013,Luitz.2015,Nicolas_MBL2019,roy.2021}.

%%%%%%%%%===========================================
\section{Model}
%{\it Model.---} 
We study the following standard  model~\cite{Oganesyan.2007,Znidaric.2008,Pal.2010,Bauer.2013,Kjall.2014,Luitz.2015,Serbyn.2015} of MBL for a fermionic chain with  an i.i.d. random onsite potential $\epsilon_i\in [-W,W]$ of strength $W$ on $i=1,\dots,L$ sites and nearest-neighbour repulsion ($V$),
\begin{equation}
\mathcal{H}=t\sum \limits_{i=1}^{L-1} \left(c_i^\dagger c_{i+1}^{\pd}+c_{i+1}^\dagger c_i^{\pd} \right)+\sum \limits_{i=1}^L \epsilon_i^{\pd} \hat{n}_i^{\pd} +V \sum \limits_{i=1}^{L-1} \hat{n}_i^{\pd}\hat{n}_{i+1}^{\pd}.
\label{Eq:many_body_Ham}
\end{equation}
Here $c_i^\dagger$ ($c_i^{\pd}$) is the fermion creation (annihilation) operator for site $i$, with number operator
$\hat{n}_i=c_i^\dagger c_i^{\pd}$. We choose $t=0.5$ and $V=1$ to be consistent with earlier studies; and
study the model in the half-filled sector at `infinite temperature', which corresponds to the middle of the many-body energy spectrum. In this case, 
%for periodic boundary conditions
the model shows a thermal to MBL transition at a critical disorder $W_c\simeq 3.7 - 4.2$~~\cite{Luitz.2015}. In this work, we take the critical disorder $W_c$ as 3.75. Variation of $W_{c}$ between $\sim 3.5$$-$$4$ leads to comparably good scaling collapse in our finite-size scaling analysis.

To describe the many-body system in Fock space, we employ the occupation-number
basis $\{\ket{I}\}$ of particles on the real-space sites, $\ket{I}= \ket{n_1^{(I)} n_2^{(I)}.. n_L^{(I)}}$ with $n_i^{(I)}\in 0$ or 1. In this basis, the Hamiltonian Eq.~\eqref{Eq:many_body_Ham} takes the form of a  tight-binding model~\cite{Welsh.2018,Logan.2019,Ghosh.2019}
\begin{equation}
\mt{H}=\sum\limits_{I,J} T_{IJ}^{\pd}\ket{I}\bra{J} +\sum \limits_I \mt{E}_I^{\pd} \ket{I}\bra{I},
\label{eq:fham}
\end{equation}
but on the FS lattice [Fig.~\ref{Fig:Fig1}(a)].  Here, the FS ``hopping'' $T_{IJ}=t$ when $\ket{I}$ and
$\ket{J}$ are connected by a single nearest-neighbor hop in real-space, and $T_{IJ}=0$ otherwise. The onsite 
``disorder'' potential at FS site $I$ is $\mt{E}_I=\sum\limits_i \epsilon_i n^{(I)}_i+ V\sum\limits_i n_{i}^{(I)} n_{i+1}^{(I)}$, and is a combination of the real-space disorder potential and the nearest-neighbor interaction.

The disorder-averaged many-body density of states 
%(DOS)
for the model is a Gaussian as a function of energy 
$E$, with the mean $\propto L$ and variance $\mu_E^2=(t^2+W^2/6+V^2/8)L/2$ \cite{Welsh.2018}. As a result, to treat different system sizes on the same footing and for the theory to have a well-defined thermodynamic 
limit~\cite{roy2020fock}, we scale the parameters and work with a rescaled $\tilde{\mt{H}}=\mt{H}/\sqrt{L}$~\cite{Welsh.2018,Logan.2019,roy2020fock}. This leads directly to the scaled version of the self-energy mentioned above.  We also set the mean many-body energy to zero, by transforming $\tilde{\mt{H}}$ to $\tilde{\mt{H}}-\mathds{1}\Tr\tilde{\mt{H}}/\mt{N}_F$ for each disorder realization, since the middle of the spectrum fluctuates with disorder realization for a finite system. 

%%%%%%%%%=============================================================
\section{Recursive Green's function method}
\begin{figure}
\includegraphics[width=1.0\linewidth]{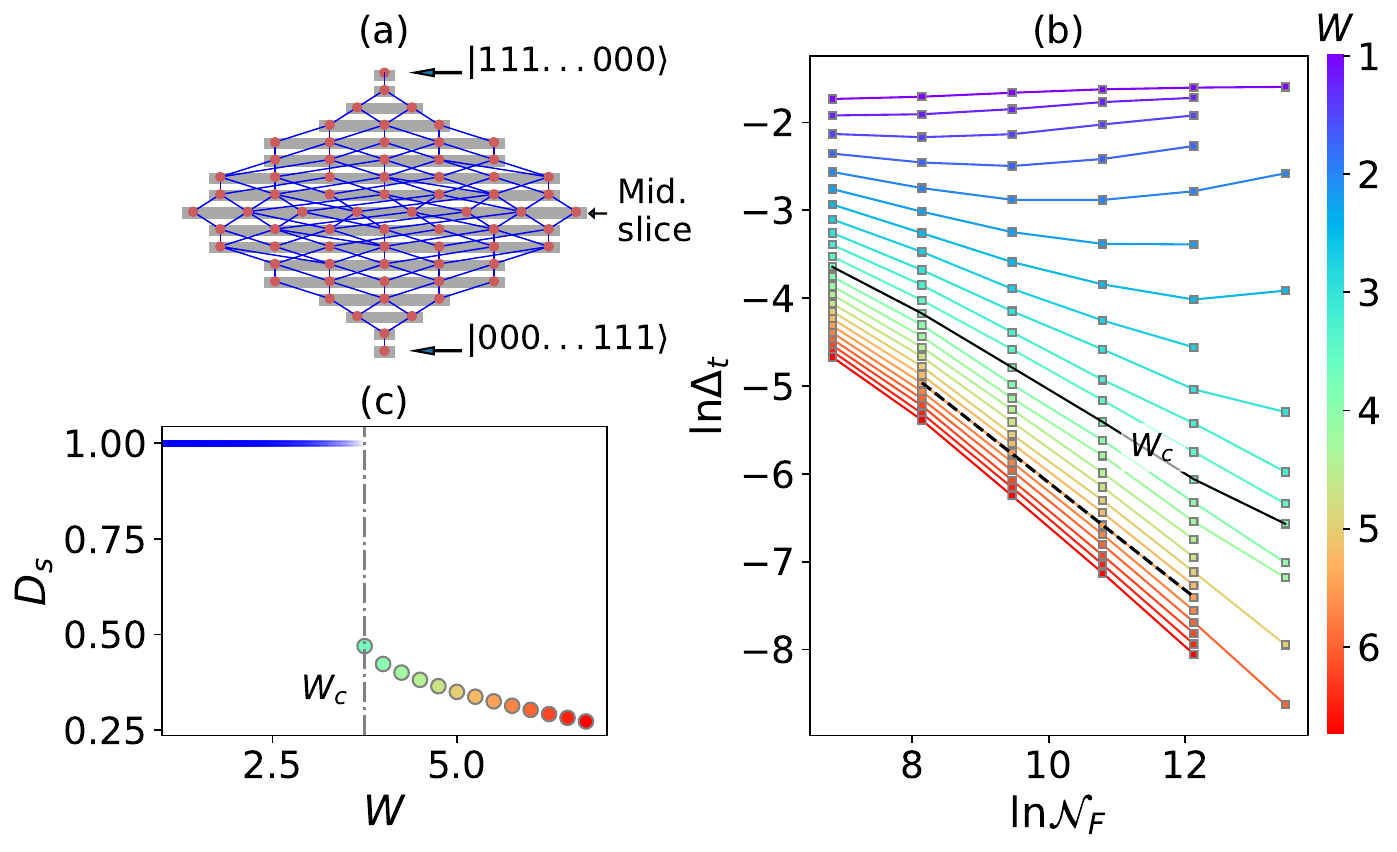}
\caption{(a) Fock-space lattice constructed out of real-space occupation-number basis states (orange circles),
illustrated 
for $L=8$, starting at the top with $\ket{111..000}$, i.e. all particles on the left side, and ending at the bottom with all particles on the right. The hoppings (blue lines) and the slices (grey lines) are indicated. (b) $\ln\dt$ as a function of $\ln\NF \propto L$ for different $W$ (color bar) across the MBL transition ($W_c$). The exponential decrease of $\dt$ with $L$ for $W>W_c$ is shown by the dashed black line (linear fit) for one value of $W$. We also show the data for $L=22$ for several disorder strengths. (c) The fractal dimension $D_s$ 
%extracted
obtained from the finite-size scaling theory jumps discontinuously across the transition, from $D_s<1$ in the MBL phase to $D_s= 1$ in the thermal phase. At $W=W_c$ (vertical dash-dotted line) 
$D_s\simeq 0.5$. }
\label{Fig:Fig1}
\end{figure}

%{\it Recursive Green's function method.---}
Elements of the retarded FS propagator $G(E)=[(E+i\eta)\mathds{1}-\tilde{\mt{H}}]^{-1}$ are computed by implementing the standard recursive method \cite{Lee.1981,MacKinnonKramer.1983,MacKinnon.1980,Verges1999,Prabhakar2021} at the middle ($E=0$) of the energy spectrum. We choose the broadening 
(or regulator)  $\eta(W)=\sqrt{2\pi}\mu_E/(\sqrt{L}\mt{N}_{F})$~\cite{Welsh.2018,Logan.2019},
the mean many-body level spacing of $\tilde{\mt{H}}$, which depends on disorder strength $W$ through the variance $\mu_E^2$. We arrange the Fock space basis states in a layered lattice structure, as illustrated in Fig.~\ref{Fig:Fig1}(a). Crucially, this is local, in that sites belonging to any slice or layer are connected through hopping only to sites belonging to the nearest-neighbor layers. It is this locality that allows for an efficient implementation of the method for the FS lattice, as detailed in Appendix~\ref{Appndx:RecursiveMethod}.

%%%%%%%%%
%%%%%%%%
 
%%%%%%%%%=============================================================
% \section{Results}

We calculate the diagonal elements $G_{II}(E)=\langle I\vert G(E)\vert I\rangle$, where $I\in$ the middle slice $M$ [Fig.~\ref{Fig:Fig1}(a)]. We extract one of the important characterizations of the localization properties in the FS, the imaginary part $\Delta_I$ of the Feenberg self-energy 
$\Sigma_I(E)=X_I(E)-i\Delta_I(E)\equiv G_{II}^{-1}(E)-(E+i\eta-\mathcal{E}_I)$ at a FS site $I$. As mentioned in the introduction, the analogous quantity on the real-space lattice was quintessential in the development of the concept of localization \cite{Anderson1958} for non-interacting systems. We consider the distributions of $\Delta_I$,  in particular its disorder-averaged typical value, denoted by $\Delta_{t}$. The latter is the geometric mean calculated by averaging over different disorder realizations and different Fock space sites using 
$\ln\dt=\braket{\ln\di}_{I,\{\epsilon_i\}}$, where  $\braket{..}_{I,\{\epsilon_i\}}$ denotes averaging over $I\in M$ and disorder realizations $\{\epsilon_i\}$. Depending on $L$, $\sim$150 to 10,000 disorder realizations 
are employed to generate the distribution of $\di$ (Appendix~\ref{Appndx:convergence}). We have checked that our results converge with number of disorder samples for all the system sizes, $L=12-22$, that we study, and the details are given in Appendix~\ref{Appndx:convergence}.

%%%%%%%%%%%%--------------------------------------
\section{Multifractality and scaling theory from the imaginary part of Feenberg self-energy}
%{\it Multifractality and scaling theory from the imaginary part of Feenberg self-energy.---}
The imaginary part $\Delta_I(E)$ of the self-energy determines the energy-resolved decay time [$\Delta_I(E)^{-1}$] of a localized initial state $\ket{I}$ or, alternatively, the life time of an excitation with energy $E$ created at the FS lattice site $I$ \cite{Anderson1958}. As such, $\Delta_I(E)$ directly encodes information about localization or lack thereof. In the thermal phase, the typical value $\dt\sim\mt{O}(1)$, as the initial state decays in a finite time, whereas in the MBL phase, $\dt\to 0$ in the thermodynamic limit $\mathcal{N}_F\to\infty$.

Numerical results for $\Delta_t$ as a  function of $\NF$ are shown in Fig.~\ref{Fig:Fig1}(b). $\dt$ indeed decreases as a function of $W$ from an $\mathcal{O}(1)$ value in the thermal phase to a value which in the MBL phase vanishes exponentially rapidly with $L$. Deep in the thermal phase at weak disorder, $\dt$ is independent of $\NF$.  
By contrast, in the MBL phase ($W>W_c$), $\dt$ decays as a power-law, $\propto\NF^{-(1-D_s)}$ with $0<D_s<1$.  
As discussed later, $D_{s}$ is a spectral fractal dimension which characterizes the multifractality of the MBL states. It depends on $W$ as shown in Fig.~\ref{Fig:Fig1}(c), where $D_s$ has been obtained from the finite-size scaling analysis discussed below. Note that at intermediate disorder in the ETH phase, 
%closer to the MBL transition, 
$\dt$ initially decays with $\NF$ for small $L$, before showing an increasing trend, presumably towards its finite asymptotic value in the thermodynamic limit [Fig.~\ref{Fig:Fig1}(b)]. This indicates that the systems are in the critical regime at intermediate disorder.

The bare data itself conforms to the expectation that in the thermodynamic limit $\dt$ vanishes in the MBL phase and approaches a finite $\mt{O}(1)$ value in the thermal phase. But to analyze compellingly the critical properties of $\dt$, we perform a scaling collapse of the data using the following finite-size scaling ansatz ~\cite{Garc_PRL2017,Nicolas_MBL2019,roy.2021},
%%%%%%%%%%%%%%%%%%
\begin{equation}
\ln\dfrac{\dt}{\Delta_c}=
\begin{cases}
      \Fv\left(\dfrac{\NF}{\Lambda}\right) & ~~:~ W<W_c\\
      \Fl\left(\dfrac{\ln\NF}{\xi}\right) & ~~:~ W>W_c,
    \end{cases}
    \label{Eq:scaling_fn_delta} 
\end{equation}
%%%%%%%%%%%%%%%%%%%
where  $\Delta_c=\dt(W$$=$$W_c)\sim \NF^{-(1-D_c)}$. 
The scaling ansatz states that in the ETH phase, $\dt$ follows a `volumic' scaling form where the finite-size scaling is controlled by a Fock-space volume scale, $\Lambda$ \cite{Nicolas_MBL2019,roy.2021,Garc_PRL2017}. In the MBL phase, on the other hand, the scaling form is `linear' with the scaling controlled by a Fock-space lengthscale, $\xi$ \cite{Garc_PRL2017,Nicolas_MBL2019,roy.2021}. Taking $W_{c}=3.75$, the good collapse of the data shown in Fig.~\ref{Fig:delta_scaling} suggests that the above scaling forms are appropriate. As shown in Appendix~\ref{Appndx:scaling_Dt}, varying $W_{c}$  between $\sim 3.5$$-$$4$ leads to comparably good scaling.

%%%%%%%%%%%
\begin{figure}
\centering
\includegraphics[width=0.9\linewidth]{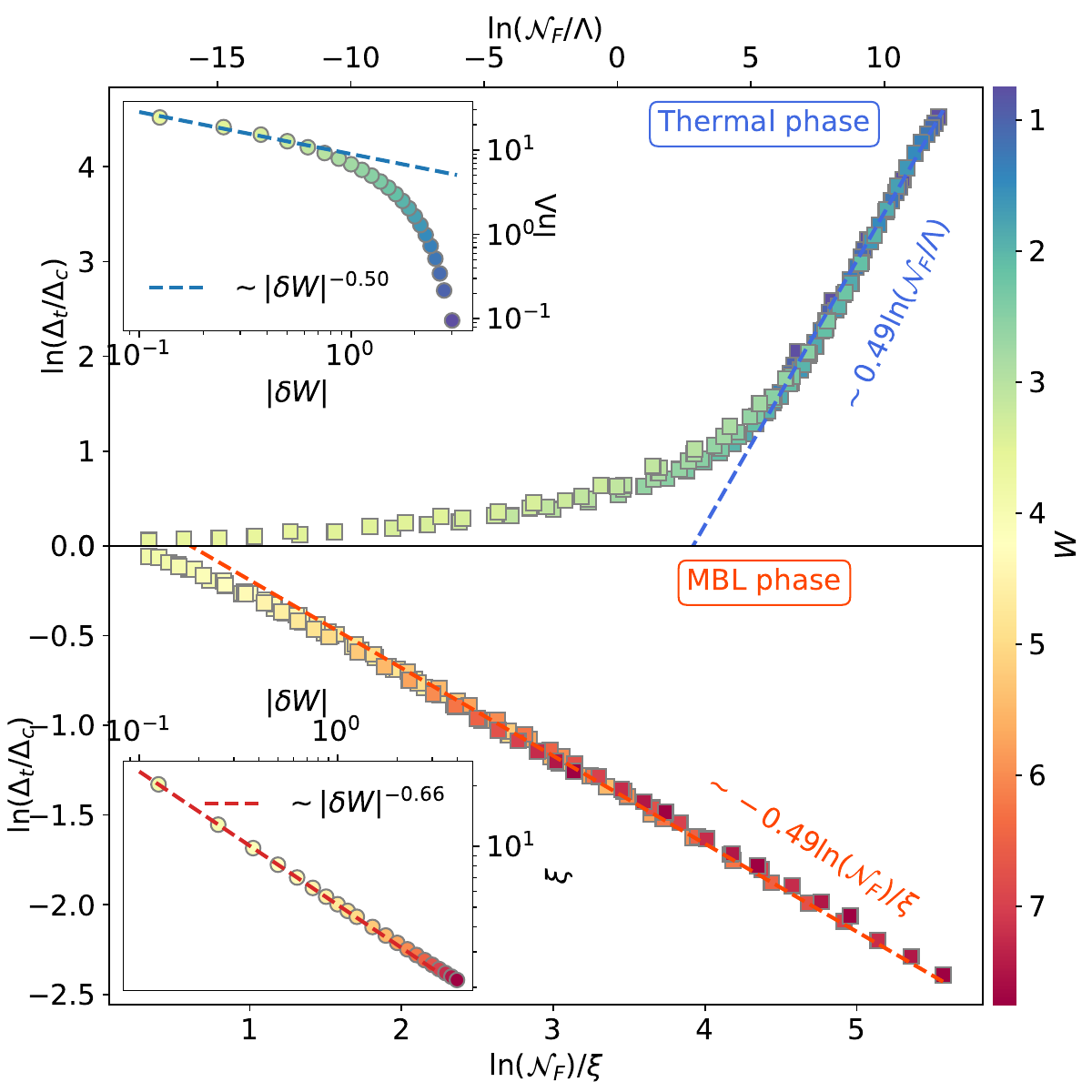}
\caption{Finite-size scaling collapse of $\ln(\dt/\Delta_c)$ using respectively the volumic and linear scaling [Eq.~\eqref{Eq:scaling_fn_delta}] in the thermal (upper panel) and MBL (lower panel) phase, with $W_c=3.75$.  
\emph{Upper panel}: The asymptotic scaling of $\ln(\dt/\Delta_c)\sim 0.5\ln(\NF/\Lm)$ deep in the thermal phase is determined by the exponent $(1-D_c)\simeq 0.5$ at $W=W_c$, where $\Delta_c\sim \NF^{-(1-D_c)}$. Inset shows the KT-like essential singularity of the non-ergodic volume, $\Lm\sim \exp{(b/\sqrt{\delta W})}$ with $\delta W=(W_c-W)$,
near $W_c$. 
\emph{Lower panel}: Deep in the MBL phase, asymptotic scaling gives $\ln(\dt/\Delta_c)\propto (\ln\NF)/\xi$. The inset shows the divergence of the correlation length, $\xi\sim |\delta W|^{-\beta}$ with $\beta \simeq 0.7$, on approaching  the transition from the MBL phase.  System sizes from $L=12-20$ were used to obtain the scaling.
}
\label{Fig:delta_scaling}
\end{figure}
%%%%%%%%%%%%%

The `non-ergodic' volume $\Lambda(W)$ \cite{Garc_PRL2017}  extracted from the scaling collapse of $\ln (\dt/\Delta_c)$ in the thermal phase ($W<W_c$) is shown in the upper panel of Fig.~\ref{Fig:delta_scaling} (inset). $\Lambda$ diverges at the critical point with a KT-like essential singularity, $\Lambda\sim \exp{[b/(\delta W)^\alpha]}$ with $\alpha\simeq 0.5$, where $\delta W=(W_c-W)$ and $b\sim \mathcal{O}(1)$. This kind of KT-type singularity has been predicted by a phenomenological RG theory \cite{Goremykina_PRL2019,Dumitrescu_KTMBL2019,morningstar2019rg,Morningstar_PRB2020}, albeit based on a real-space picture. Throughout the ETH phase, since the eigenstates are understood to be 
ergodic~\cite{Nicolas_MBL2019}, we expect $\dt\sim\mt{O}(1)$ asymptotically in the limit $\NF\gg\Lambda$. This implies for the volumic scaling function in Eq.~\eqref{Eq:scaling_fn_delta} that $\mt{F}_\mathrm{vol}(x)\sim (1-D_c)\ln x$ for $x \gg 1$. As evinced in the upper panel of Fig.~\ref{Fig:delta_scaling}, this is indeed the asymptotic form of the scaling function, $\mt{F}_\mathrm{vol}(x)\sim 0.49 \ln  x$ (see Appendix~\ref{Appndx:scaling_Dt} for details) from which we estimate $D_c\simeq 0.51$. This is an excellent agreement with $D_c\simeq 0.5$ obtained by fitting the raw data of $\dt$ against $\NF$ for $W_{c}=3.75$ [Fig.~\ref{Fig:Fig1}(b)]. A significant conclusion from this is that $D_s=1$ throughout the ETH phase and jumps \emph{discontinuously} to a value $D_s<1$ at the MBL transition, whereafter it decreases smoothly with increasing $W$ [Fig.~\ref{Fig:Fig1}(c)]. This is in complete consonance with the discontinuity across the MBL transition of the fractal dimension characterising the Fock-space IPRs of the eigenstates~\cite{Nicolas_MBL2019,roy.2021}.
Furthermore, the form of the divergence of $\Lambda$ and the asymptotic form of $\mathcal{F}_\mathrm{vol}$ implies that in the thermodynamic limit, $\Delta_t$ vanishes continuously as the transition is approached from the ETH side as $\Delta_t\sim \Lambda^{-(1-D_{c})}\sim \exp[-b(1-D_c)/\delta W^\alpha]$.

In the MBL phase for $W>W_c$, the scaling collapse yields a diverging correlation length,  $\xi\sim (W-W_c)^{-\beta}$ with $\beta \simeq 0.7$,  as shown in the lower panel of  Fig.~\ref{Fig:delta_scaling} (inset). For $x=(\ln\NF)/\xi\gg 1$, the asymptotic scaling is $\mt{F}_{\text{lin}}(x)\sim -(1-D_c) x$ (Appendix~\ref{Appndx:scaling_Dt}), 
as indeed seen in Fig.~\ref{Fig:delta_scaling} lower panel. From the form of $\Fl(x)$, $\xi = (1-D_c)/(D_c-D_s)$, which implies a divergent $\xi$ as $D_s\to D_c$, i.e.\ $W\to W_{c}+$ (Appendix~\ref{Appndx:scaling_Dt}).
  Note that $D_c<1$ implies that the MBL critical point is actually a part of the MBL phase itself. This is consistent with the understanding that the entire MBL phase is critical in the sense that it is multifractal, and the MBL transition can be understood as the terminal end point of the line of fixed points.

Having established the critical scaling of the FS order parameter $\dt$ and the volume (length) scale in the thermal 
(MBL) phase, we now discuss how the system-size scaling $\dt\sim \NF^{-(1-D_s)}$ in Fig.~\ref{Fig:Fig1}(b),(c) for $W>W_c$ is related to multifractality in the MBL phase. MBL eigenstates are known to be multifractal in 
nature~\cite{deluca2013,Luitz.2015,Nicolas_MBL2019,roy.2021,DeTomasiRare2021}, i.e. non-ergodic but extended over 
$\sim\NF^D$ ($0<D<1$) FS sites. Similar to the case of the local density of states discussed in Ref.~[\onlinecite{Altshuler2016a}], the multifractality can be deduced from the dependence of the typical value and the distribution of $\Delta_I$, on the broadening parameter $\eta$ for large but finite $\mathcal{N}_F$. 
In this limit, $\Delta_t$ saturates as a function of $\eta$ below an energy scale $\eta_c\sim \mathcal{N}_F^{-z}$ ($0<z<1$), namely, $\dt\sim \eta_c^\theta\sim \mathcal{N}_F^{-(1-D_s)}$ for $\eta\ll \eta_c$, whereas $\Delta_t\sim \eta^\theta$ for $\eta\gg \eta_c$. Here the exponent $\theta>0$ and the spectral fractal dimension $D_s=1-z\theta$ lies between 0 and 1~\cite{Altshuler2016a}. Physically, $\eta_c$ signifies the presence of an energy scale much larger than the mean level spacing $\NF^{-1}$ for multifractal states for any finite system. In our calculations, $\eta\propto \mathcal{N}_F^{-1}\ll \eta_c$, thus we expect $\Delta_t\sim \mathcal{N}_F^{-(1-D_s)}$. This is indeed the behaviour for $\Delta_t(\mathcal{N}_F)$ that we find in Fig.~\ref{Fig:Fig1}(b) over the entire MBL range ($W>W_c$).
It is also completely consistent with the asymptotic form of the scaling function [Fig.~\ref{Fig:delta_scaling} (lower panel)] and the $D_s$ extracted from it [Fig.~\ref{Fig:Fig1}(b)]. The spectral dimension $D_s$ can be shown to be the same as the fractal dimension $D$ extracted from the eigenstates under very general considerations \cite{Altshuler2016a}. We note that the behaviour $\dt\sim \NF^{-(1-D_s)}$ is different from a self-consistent theory~\cite{Logan.2019,roy2020fock} of MBL, where  the thermodynamic limit of $\NF\to\infty$ is taken first before taking the $\eta\to 0$ limit.  As a result, one obtains $\Delta_t\propto \eta$ and such a theory does not in effect distinguish between Anderson localized and multifractal states.

\begin{figure}
\includegraphics[width=1.0\linewidth]{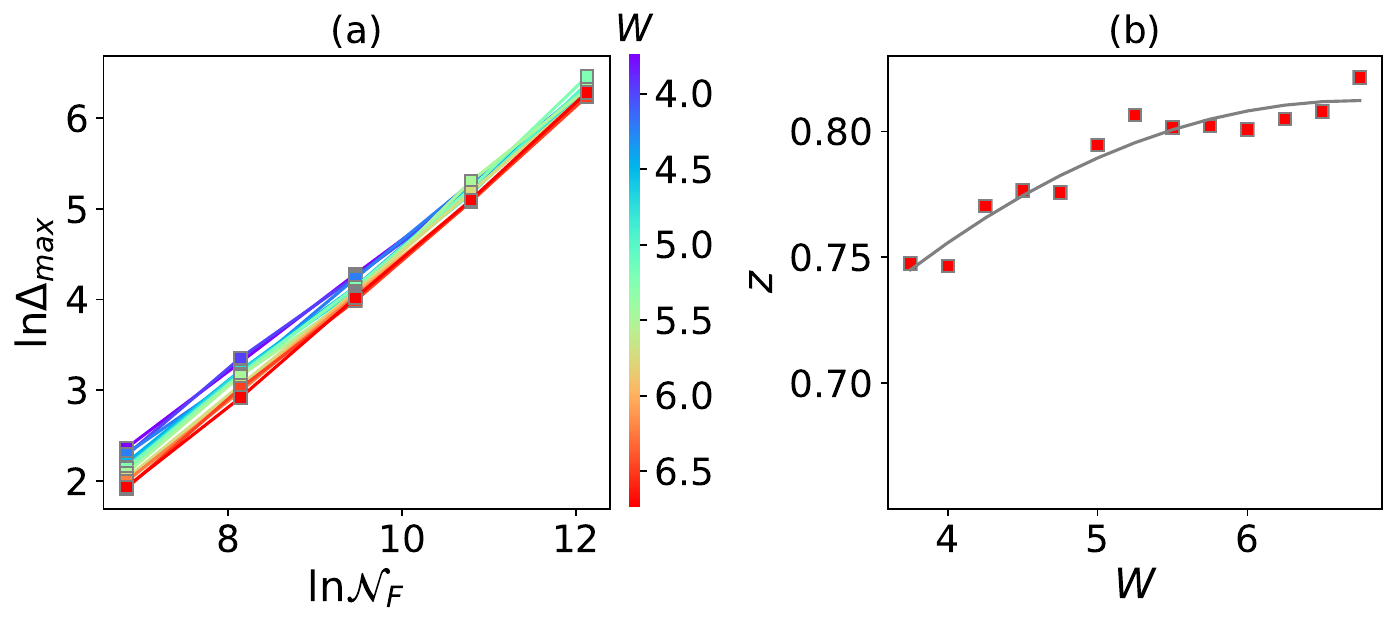}
\caption{(a) In the MBL phase, the  scaling of $\Delta_\mathrm{max}$ with the Fock space dimension $\NF$ is shown to follow a power-law $\Delta_{\mathrm{max}}\propto\NF^{z}$ for different disorder strengths (colorbar). To obtain good statistics, $\ln\Delta_{\mathrm{max}}$ plotted here is averaged over a few 
of the largest values of $\ln\Delta$ for a given $L$ and $W$.
(b) The exponent $z$, obtained from $\Delta_\mathrm{max}\sim \eta_c^{-1} \sim \NF^z$ in (a), is shown to be $<1$ in the MBL phase, consistent with the multifractal scaling of the self-energies. The solid line is merely a guide to the eye.
}
\label{fig:dists-collapse}
\end{figure}

Further insight into the multifractal behaviour may be gained by analysing the tail of the probability distribution
function $P(\Delta)$ of $\Delta_I$ over disorder realizations. For finite $\eta$ and/or $\NF$, one expects 
\cite{Altshuler2016a} the distribution to be cut off at a maximum value
$\Delta=\Delta_\mathrm{max}$, where $\Delta_\mathrm{max}\sim 1/\eta_c\sim \NF^z$ for $\eta\ll \eta_c$. As shown in Fig.~\ref{fig:dists-collapse}(a), we indeed find that $\Delta_\mathrm{max}$ directly extracted from the numerical data follows a power-law $\sim \NF^{z}$. The exponent $z$ extracted from $\Delta_\mathrm{max}$ is shown in 
Fig.~\ref{fig:dists-collapse}(b), and indeed satisfies $0<z<1$. This further confirms the multifractal scaling of the self-energy in the MBL phase.

%%%%%%%%%%%%--------------------------------------
\section{Conclusion}
%{\it Conclusion.---} 
Using a recursive Green function method, we have in summary obtained the critical scaling across the MBL transition
by calculating the self-energy associated with the local Fock-space propagator, the typical value  of the imaginary part of which, $\Delta_{t}$, acts as an order parameter for the MBL transition.
The finite-size scaling of  $\Delta_{t}$ implies the existence of a non-ergodic volume with a KT-like essential singularity, which directly controls the continuous vanishing of $\Delta_{t}$ on approach to the transition from the ergodic side; and the multifractal nature of the MBL phase was demonstrated via determination of the spectral fractal dimension $D_{s}$, which was shown to change discontinuously across the transition. While our focus here has been on the local FS propagator, the recursive Green function  method also gives access to the non-local propagator,  a question of immediate future interest which can potentially provide insights into the critical scaling of a FS localization length, and also enable study of the inhomogeneous nature of MBL eigenstates on the FS 
lattice~\cite{roy.2021}. It would also be interesting to explore further the connection \cite{roy.2021} between 
real-space and Fock-space critical properties,  e.g. how rare thermal regions in real space~\cite{DeRoeck2017,Luitz2017,Potirniche2019} affect the FS self-energy. In the same vein, possible connections between the Fock-space propagators and one-particle density matrix and real-space propagators \cite{Bera2015,Lezama2017,Hopjan2020,Hopjan2021,Orito2021,Jana2021} also remains a question for future work.

\begin{acknowledgements}
SB acknowledges support from SERB (ECR/2018/001742), DST, India. SM acknowledges support from QuST, DST, India. SR acknowledges support from an ICTS-Simons Early Career Faculty Fellowship via a grant from the Simons Foundation (677895, R.G.) and EPSRC Grant No. EP/S020527/1.
\end{acknowledgements}
%%%%%%%%%======================================================================
%%%%%%%%%======================================================================
%\section*{Appendices}
\appendix
\renewcommand{\thefigure}{A\arabic{figure}}
\setcounter{figure}{0}

\section{Recursive Green function Method}
\label{Appndx:RecursiveMethod}
In the recursive method, we calculate different elements of the Green function by starting from one of the two minimally connected Fock-space sites ($\ket{111\dots 000}$, $\ket{000\dots111}$) and by adding the next slice connected by hopping to the earlier slice at each step. Here we start with the top site $\ket{111...000}$ as shown in Fig.~\ref{Fig:Fig1}(a). At each iteration step, the method inverts a matrix containing the Hamiltonian elements for the added slice and is, therefore, of size $\mt{N}_\ell\times \mt{N}_\ell$, where $\mt{N}_\ell$ is the size of the added slice. Hence, this method avoids the inversion of the full Hamiltonian in Eq.~\eqref{eq:fham}. The maximum size of the matrix ($\mt{N}_M\times \mt{N}_M$) that is inverted during the calculation is determined by the number of sites $\mt{N}_M$ in the middle slice, which is the largest slice of the Fock space [Fig.~\ref{Fig:Fig1}(a)]. 
For $L$ between $10$ and $22$, $\mt{N}_{M}$ is between $\sim 1-2$ orders of magnitude lower than the full Fock-space
dimension $\mathcal{N}_{F} =\binom{L}{L/2}$. It is this substantial reduction that provides the advantage to achieve the system size $L=22$ even with serial computation. The method can be applied to even larger systems like $L=24$ with parallelization.

\begin{figure*}
    \centering
    \includegraphics[width=0.7\linewidth]{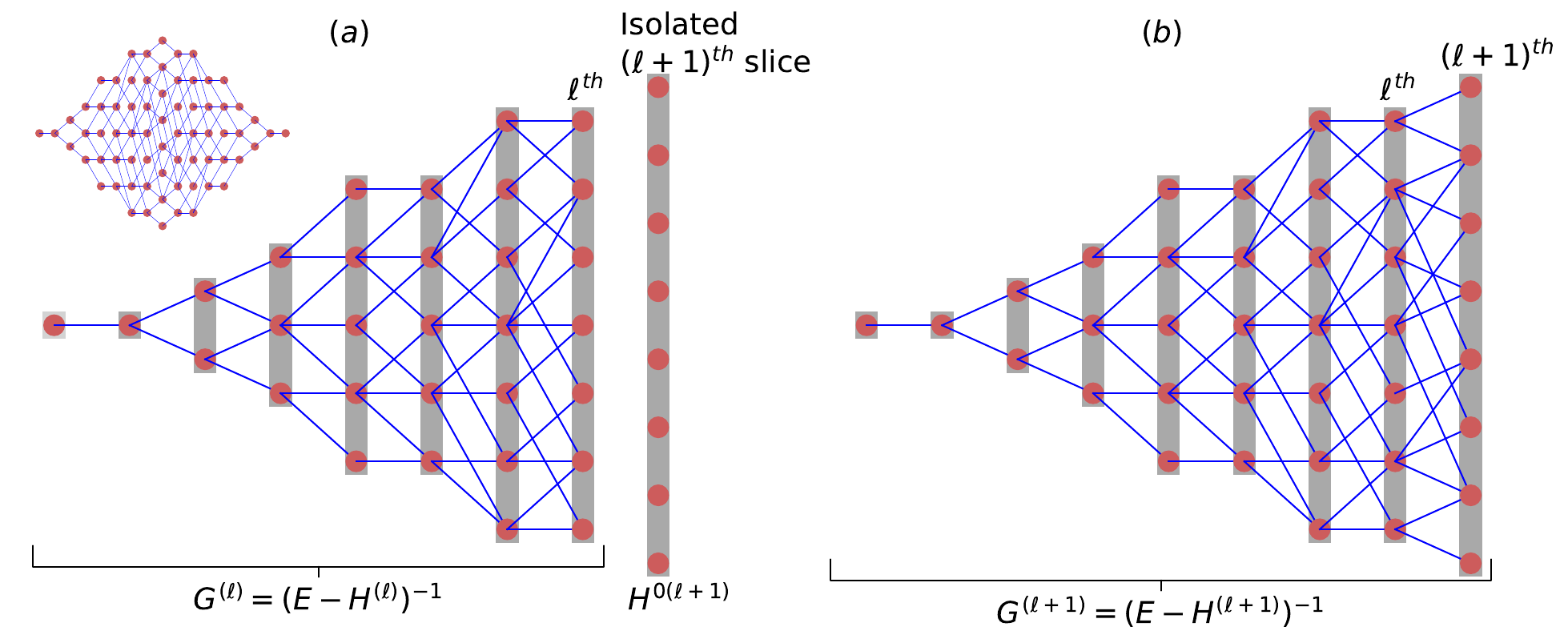}
    \caption{(a) The figure on the top left corner is similar to Fig.~\ref{Fig:Fig1}(a), but rotated by $90^{\circ}$. The larger figure shows a part of it, up to $\ell$th slice, where the grey shaded area shows each slice. Orange circles and solid blue lines represent the FS sites and hopping, respectively. (b) As we keep on adding the next slices, the Green function of the left part gets updated via the iterative Eqs.~\eqref{Eq:itrtv_Gfn_eqn}.}
    \label{SFig:recursive_Gf}
\end{figure*}

At the $\ell$-th step of recursion, let $H^{(\ell)}$ be the Hamiltonian corresponding to the part of the Fock space lattice containing all slices up to the $\ell$-th slice. The corresponding Green function for these $\ell$ slices is given by $G^{(\ell)}=\left[E^+\mathds{1}-H^{(\ell)}\right]^{-1}$, where $E^+=E+i\eta$. Using Eq.~\eqref{eq:fham} and the local structure of the Fock space lattice, the Hamiltonian corresponding to all slices up to the $(\ell+1)$-th slice can be written as:
\begin{eqnarray}\nonumber
&&H^{(\ell+1)}=\begin{pmatrix}
H^{(\ell)} & \mathcal{T}_{lr}\\
T_{rl} & H^{(0)(\ell+1)}\end{pmatrix},\\ \nonumber
&&H^{(0)(\ell+1)}=\sum \limits_{I\in (\ell+1)} \mt{E}_{I,\ell+1}\ket{I,\ell+1}\bra{I,\ell+1},\\ \nonumber
&&\mathcal{T}_{lr}=\sum\limits_{I \in \ell}\sum\limits_{J\in (\ell+1)}T_{I,J}\ket{J,\ell+1}\bra{I,\ell},
\end{eqnarray}
where $H^{(0)(\ell+1)}$ contains the part of the Hamiltonian corresponding to the $(\ell+1)$-th slice disconnected from the rest of the Fock lattice, $\mathcal{T}_{lr}$ contains the hopping elements for hopping from the $\ell$-th slice to the $(\ell+1)$-th slice and $\mathcal{T}_{rl}^{\pd}=\mathcal{T}_{lr}^\dagger$. $\ket{I,\ell}$, representing the state $\ket{I}$, explicitly mentions the slice index $\ell$ which $\ket{I}$ belongs to.
Similarly, the Green function at the $(\ell+1)$-th step can be written as 
\begin{equation}\nonumber
G^{(\ell+1)}=\begin{pmatrix}
G^{(\ell+1)}_l & G^{(\ell+1)}_{lr}\\
G^{(\ell+1)}_{rl} & G^{(\ell+1)}_r
\end{pmatrix}.
\end{equation}
Here the subscript $l$ denotes the matrix elements of the left part (first to the $\ell$-th slice), and $r$ denotes the elements of the $(\ell+1)$-th slice, as described in Fig.~\ref{SFig:recursive_Gf}. Then, at the $(\ell+1)$-th step of recursion, the Green function $G^{(\ell+1)}$ and the Hamiltonian $H^{(\ell+1)}$ have the following matrix equation.
\begin{eqnarray}\nonumber
&& \left[E^+\mathds{1}-H^{(\ell+1)}\right]G^{(\ell+1)}=\mathds{1}\\
&& \begin{pmatrix}
\big(G^{(\ell)}\big)^{-1} & \mathcal{T}_{lr}, \\
\mathcal{T}_{rl} & \big(G^{0(\ell+1)}\big)^{-1}
\end{pmatrix}
\begin{pmatrix}
G^{(\ell+1)}_{l} & G^{(\ell+1)}_{lr} \\
G^{(\ell+1)}_{rl} & G^{(\ell+1)}_{r}
\end{pmatrix}
=\mathds{1}.
\label{Eq:2by2_Gr_fn_eqn_Fs}
\end{eqnarray}
Here $\big(G^{(\ell)}\big)^{-1}=(E^+\mathds{1}-H^{(\ell)})$ and $\big(G^{0(\ell+1)}\big)^{-1}$ is given by $[E^+\mathds{1}-H^{(0)(\ell+1)}]$.

%%%%%%%%%%%
%%%%%%%%%%%%5
From the above equation, we can obtain the following equation 
\[
\begin{split}
& G^{(\ell+1)}_{l}=G^{(\ell)}+ G^{(\ell)}\mathcal{T}_{lr}G^{(\ell+1)}_{rl}, \\
 & 
\braket{s| G^{(\ell+1)}_{l}|s'}=\braket{s|G^{(\ell)}|s'} +\\
&\hspace{2cm}\braket{s| G^{(\ell)}\ \big(\mathbb{T}_{\ell}\ket{\ell}\bra{\ell+1}\big)\ G^{(\ell+1)}_{rl}|s'};~  s,s'\le \ell.
\end{split}
\]
Note that $\braket{s|A|s'}=A(s,s')$ denotes an element of a rectangular matrix $A$, as each of the slices, indexed by $s$ and $s'$, contains 
different numbers of lattice points in general. For the same reason, $\mathbb{T_\ell}$ is also a rectangular matrix containing hopping matrix elements between $\ell$-th and $(\ell+1)$-th slices. Therefore,
%\[
\begin{multline*}
G^{(\ell+1)}(s,s')=G^{(\ell)}(s,s')+ \\
\hspace{2cm} G^{(\ell)}(s,\ell)\mathbb{T}_\ell G^{(\ell+1)}(\ell+1,s'),~~ s,s'\le \ell.
\end{multline*}
%%%%%%%%%%
Similarly, starting from Eq.~\eqref{Eq:2by2_Gr_fn_eqn_Fs},
\begin{subequations}
\begin{equation}
G^{(\ell+1)}(\ell+1,s)=G^{(\ell+1)}(\ell+1,\ell+1)\mathbb{T}^\dagger_\ell G^{(\ell)}(\ell,s),~~ s\le \ell,
\end{equation}
\begin{equation}
G^{(\ell+1)}(s,\ell+1)=G^{(\ell)}(s,\ell)\mathbb{T}^\dagger_\ell G^{(\ell+1)}(\ell+1,\ell+1),~~s\le \ell.
\end{equation}
\text{Therefore, for $s,s'\le \ell$,} 
%\begin{equation}
\begin{multline}
G^{(\ell+1)}(s,s')=G^{(\ell)}(s,s')+ \\
\hspace{2cm}G^{(\ell)}(s,\ell)\mathbb{T}_\ell G^{(\ell+1)}(\ell+1,\ell+1)\mathbb{T}_\ell G^{(\ell)}(\ell,s'),
\end{multline}
%\end{equation}
\text{and on-slice elements} 
\begin{equation}
G^{(\ell+1)}(\ell+1,\ell+1)=\Big[ \big(G^{0(\ell+1)}\big)^{-1}-\underbrace{\mathbb{T}^\dagger_\ell G^{(\ell)}(\ell,\ell) \mathbb{T}_\ell}_{\text{self energy }\Sigma^{(\ell)}} \Big]^{-1}.
\end{equation}
\label{Eq:itrtv_Gfn_eqn}
\end{subequations}
%%%%%%%%%%%%
Using Eqs.~\eqref{Eq:itrtv_Gfn_eqn}, we can calculate the desired elements of the Green function, avoiding the inversion of a matrix having the dimension of the full Fock space.
%%%%=====================================================
%\section{Distribution of Feenberg %self-energy}\label{Appndx:dist_D_t}

%%%%%%%%%%%%%%
\begin{figure}
\includegraphics[width=1.0\linewidth]{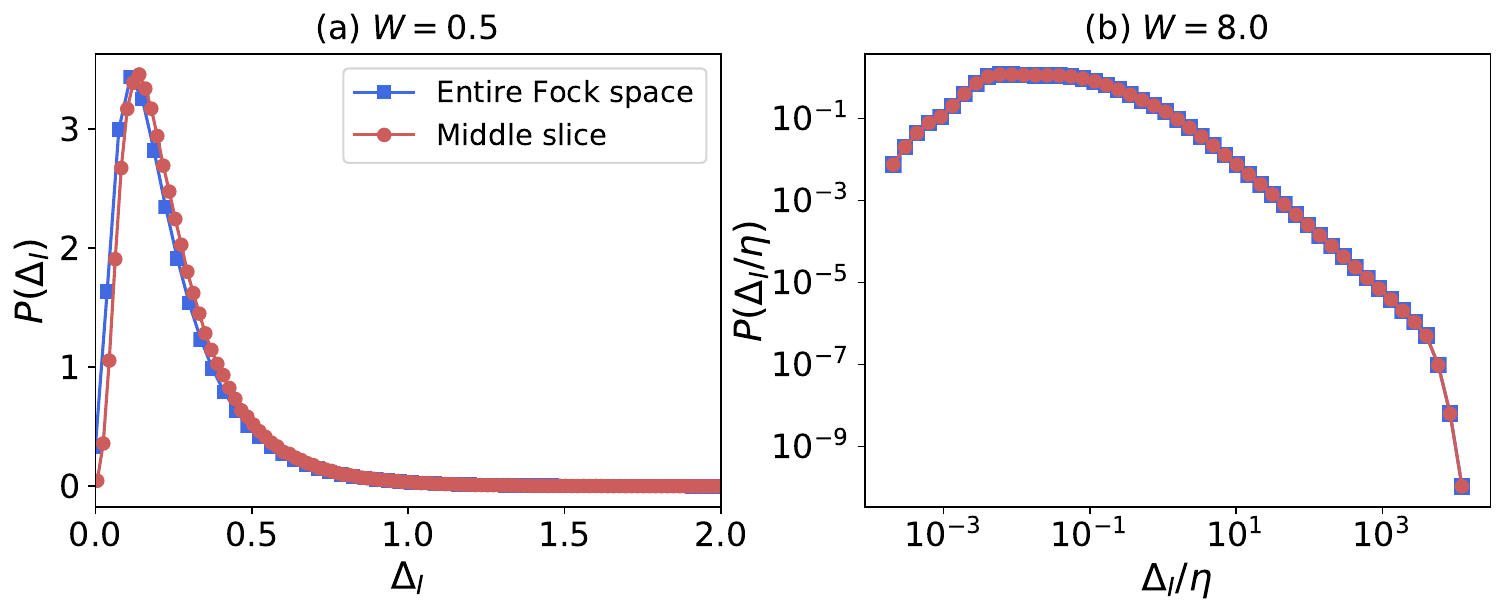}
\caption{The probability distributions of the imaginary part $\Delta_{I}$ of the self-energy,
considering both $I\in M$ and $I\in$ FS. These match well with each other, in both the thermal phase 
(shown in (a)) and  MBL phase ((b)). Data shown in the figure are for system size $L=14$. 
}
\label{SFig:dist_delta_fullFS}
\end{figure}
%%%%%%%%%%%%%%%%%

\section{Scaling of $\dt$ }\label{Appndx:scaling_Dt}
\textbf{Asymptotic behaviour of the scaling functions:}

In the ergodic phase, the scaling ansatz Eq.~\eqref{Eq:scaling_fn_delta} reads
\begin{equation}
\ln\dt=\ln\Delta_c+ \Fv(\NF/\Lm)
\nonumber
%\label{Eq:lnD_Fv}
\end{equation}
%%%%%%%%%5
with the critical $\Delta_c = C_c\NF^{-(1-D_c)}$ and $C_c$ a constant; equivalently,
\begin{equation}
\begin{split}
\Fv\left(\dfrac{\NF}{\Lm}\right) 
&=
\ln\dt- \ln C_c+(1-D_c)\ln\Lm+ \\
& \hspace{3cm}(1-D_c)\ln \dfrac{\NF}{\Lm} .
\end{split}
\label{Eq:Fv_lnD}
\end{equation}
%%%%%%%%%%%%%%
Deep in an ergodic phase, both $\Lm$ and $\dt$ are $\mathcal{O}(1)$. From Eq.~\eqref{Eq:Fv_lnD}, 
for $x=\NF/\Lm \gg 1$ the asymptotic large-$x$ behaviour of $\Fv(x)$ is then
\begin{equation}
\Fv(x)\sim (1-D_c) \ln x
\label{Eq:Algy}
\end{equation}
up to $\mathcal{O}(1)$ corrections. Its slope gives the fractal exponent $(1-D_c)$ for $W=W_{c}$, which is also 
found to be consistent with the numerical asymptotic fit in Fig.~\ref{Fig:Fig1}(b).
\begin{figure*}[t]
    \includegraphics[width=0.485\linewidth]{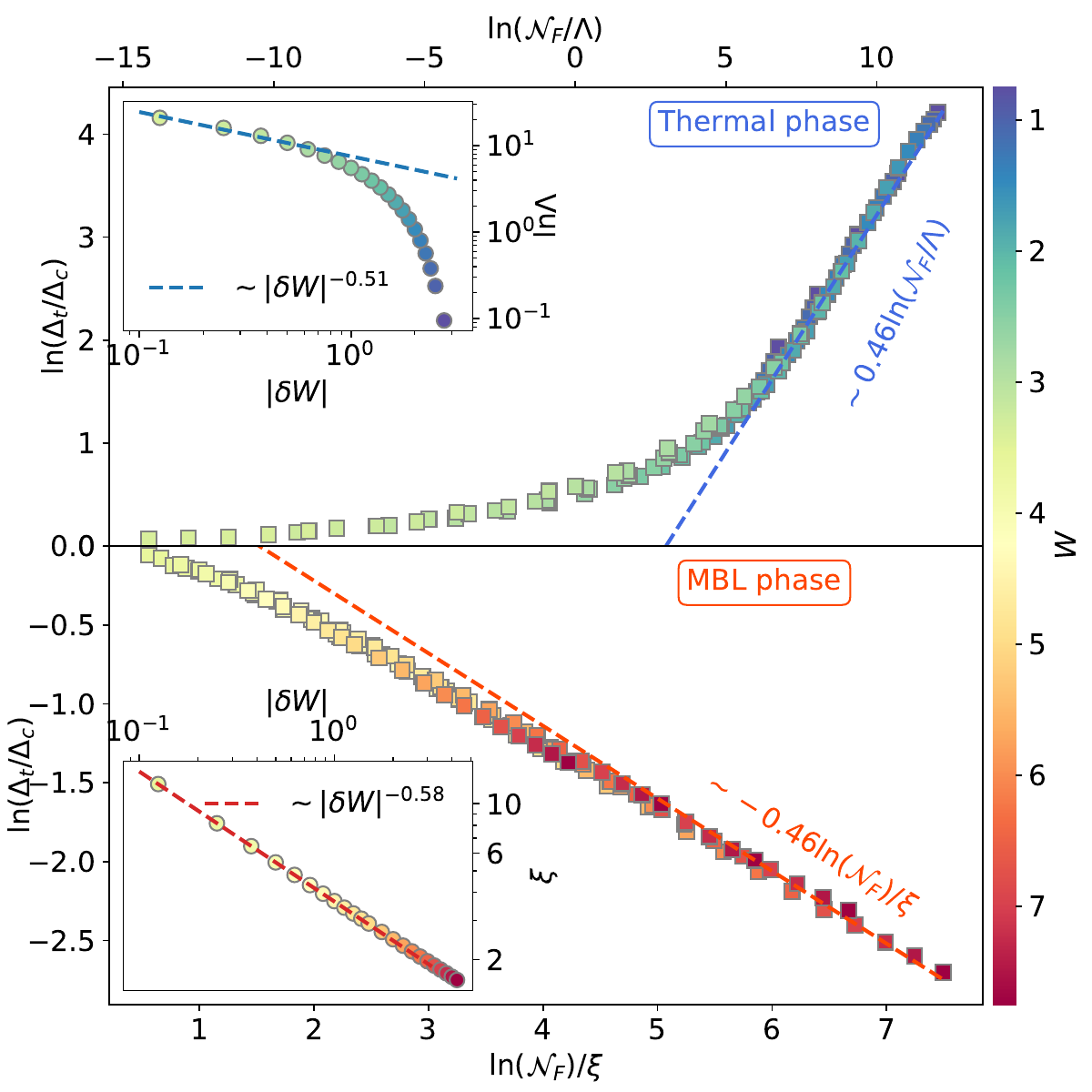}
		\includegraphics[width=0.485\linewidth]{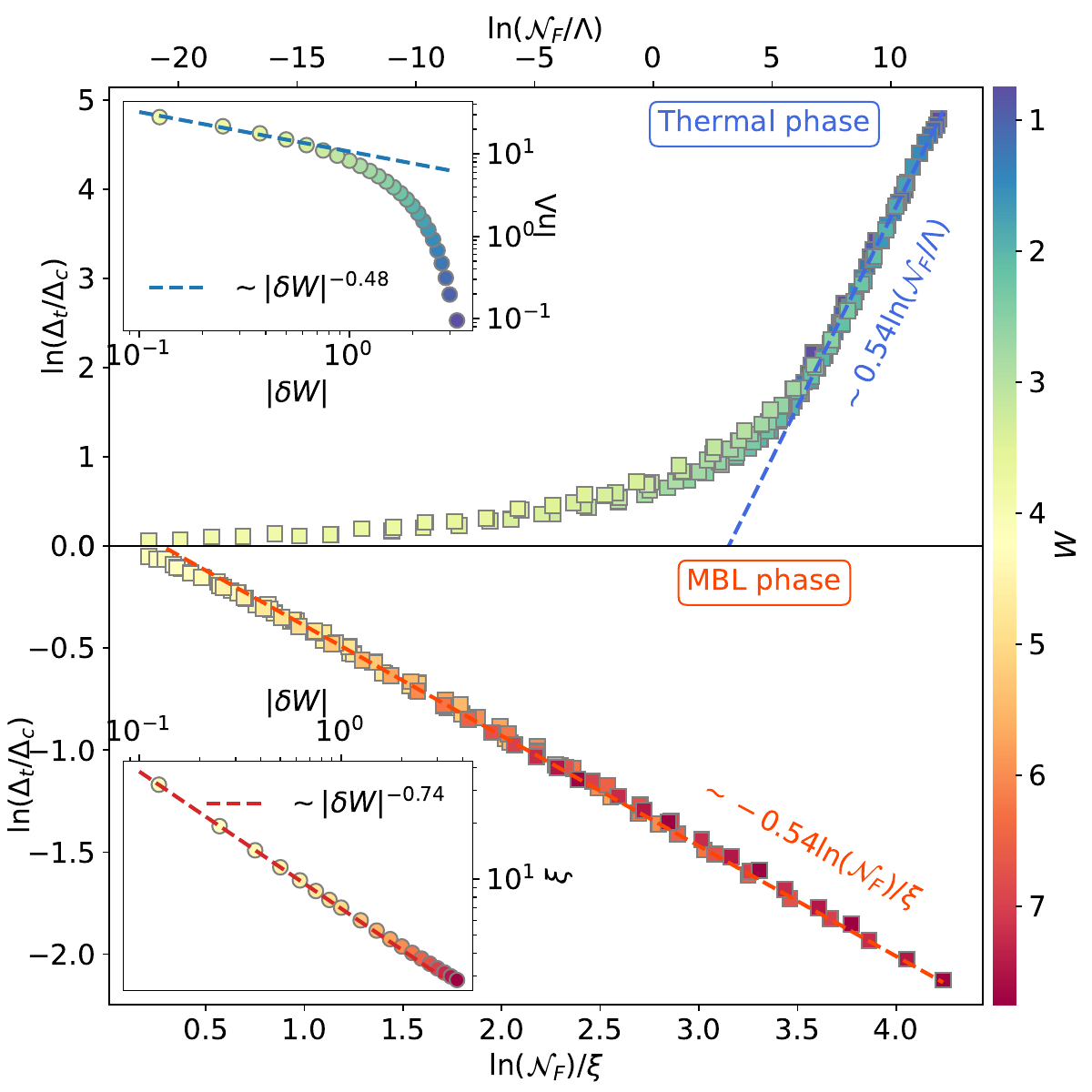}
    \caption{Finite-size scaling collapse of $\ln(\dt/\Delta_c)$ in direct parallel to that of 
		Fig.~\ref{Fig:delta_scaling} for $W_{c}=3.75$, but here shown for $W_{c}=3.5$ (left panel) and
		$W_{c}=4.0$ (right panel).
		}
    \label{SFig:LastFig}
\end{figure*}

On the other hand, in the MBL phase  $W>W_{c}$,  $\dt = C_m\NF^{-(1-D_s)}$ (Fig.~\ref{Fig:Fig1}(b)),
where $C_m$ and $D_s$ both depend on $W$. The scaling ansatz Eq.~\eqref{Eq:scaling_fn_delta} then gives
\begin{equation}
\Fl\left(\dfrac{\ln\NF}{\xi}\right)
 =
\ln\dfrac{C_m}{C_c}- 
(1-D_c) \dfrac{\ln\NF}{\xi},
\label{Eq:F_lin_lnN}
\end{equation}
%%%%%%%
with $\xi$ defined as
\begin{equation}
\xi= \dfrac{(1-D_c)}{D_c-D_s}.
\end{equation}
As $D_s\rightarrow D_c^+$, i.e. as $W\rightarrow W_c^+$, we see that $\xi$ diverges.
Deep in the MBL phase, where $x=(\ln\NF)/\xi\gg 1$, the asymptotic scaling is clearly
\begin{equation}
\Fl(x) \sim -(1-D_c) x.
\end{equation}

Finally, returning to the ergodic phase, Eq.~\eqref{Eq:Fv_lnD} with $x=\mathcal{N}_{F}/\Lambda$
can be cast as
\begin{equation}
\dt ~=~\Lambda^{-(1-D_{c})}~\exp\left[\Fv(x) -(1-D_{c})\ln x +\ln C_{c}\right]
\label{Eq:Biggles}
\end{equation}
where (as shown in the main text) the non-ergodic volume $\Lambda \sim \exp[b/\sqrt{W_{c}-W}]$ is determined by $W$, 
and $C_{c}$ is a constant. For any $W<W_{c}$, and hence for any finite $\Lambda$ no matter how large, the thermodynamic limit corresponds to $x\to \infty$. From Eq.\ \eqref{Eq:Algy} the argument of the exponential in 
Eq.\ \eqref{Eq:Biggles} is then $\mathcal{O}(1)$, so in the thermodynamic limit the order parameter $\dt$ vanishes continuously as
\begin{equation}
\dt ~\sim ~ \Lambda^{-(1-D_{c})}
\label{Eq:Bertie}
\end{equation}
and is controlled by $\Lambda$.

%%%%%%%%%%%%%%%%%%%%%%%%%%%%%

~\\
\textbf{Dependence of scaling results on $W_c$:} For the finite-size scaling shown in Fig.~\ref{Fig:delta_scaling}, $W_{c}=3.75$ was employed. 
Fig.\ \ref{SFig:LastFig} gives corresponding results for the scaling collapse on choosing
$W_{c}=3.5$ and $4$. As seen, the quality of scaling is quite robust to the variation in $W_{c}$.
%%%%===========================================
\section{Convergence of the data with the number of FS sites and disorder realizations}
\label{Appndx:convergence}\
We first verify that the statistics of the imaginary part of the self-energy ($\Delta_I$) is not affected by considering only the middle slice of the Fock space to generate the distribution. For both the ergodic and MBL phases, Fig.~\ref{SFig:dist_delta_fullFS} gives a representative example of the fact that the distributions of $\Delta_I$ ($P(\Delta_I)$ or $P(\Delta_I/\eta)$) 
are barely affected whether one considers $I\in M$ or $I\in$ all Fock-space sites.

In Table \ref{Table:disNum}, we show the number of disorder realizations considered for different sizes to generate the distributions of diagonal elements of the Green function. Due to finite computational time and resources, we follow the standard path of decreasing number of disorder realizations with increasing system size~\cite{suntajs2020quantum,Nicolas_MBL2019,Khemani.2017,Luitz.2015}. 

\begin{figure*}[t]
\centering
\includegraphics[width=\textwidth]{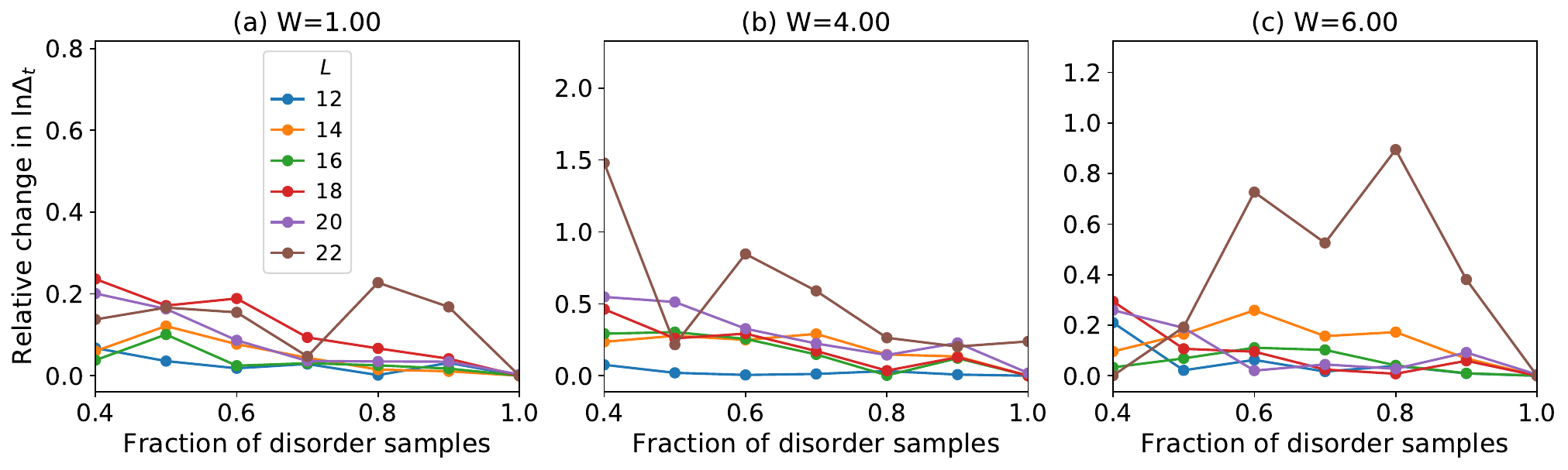}
\caption{Relative {\it percentage} change in $\mathrm{ln}\Delta_{t}$ as the sample size is changed from 0.4 times the maximum sample size to the maximum sample size. The relative change is defined as $\left|(\mathrm{ln}\Delta_{t}-\mathrm{ln}\Delta_{t}^{(1)})/\mathrm{ln}\Delta_{t}^{(1)}\right|\time 100$, where $\mathrm{ln}\Delta_{t}^{(1)}$ is calculated considering all the available samples. This shows the value of $\mathrm{ln}\Delta_{t}$ is always within $1\%$ of the values  plotted in Fig.~\ref{Fig:Fig1}(b) even for smaller sample sizes, and for all system sizes up to $L=22$.}
\label{AppFig:convergence-rel}
\end{figure*}

In Fig.~\ref{AppFig:convergence-rel}, 
%\blue{SR:I have changed the order in which the figures A4 and A5 appear to keep it consistent with the order in which they were referred to in the text} 
we show the relative change in the same quantity with respect to the value of $\ln\dt$ for the largest sample size for all system sizes. The relative change is always less than 1\% of the values plotted in Fig.~\ref{Fig:Fig1}(b), even for smaller sample sizes for all system sizes up to $L = 22$.
%%%%%%%%%%%%%%%%%%%%%%%%%%%

%%%%%%%%%%%%%
\begin{figure*}[h!]
\centering
\includegraphics[width=\textwidth]{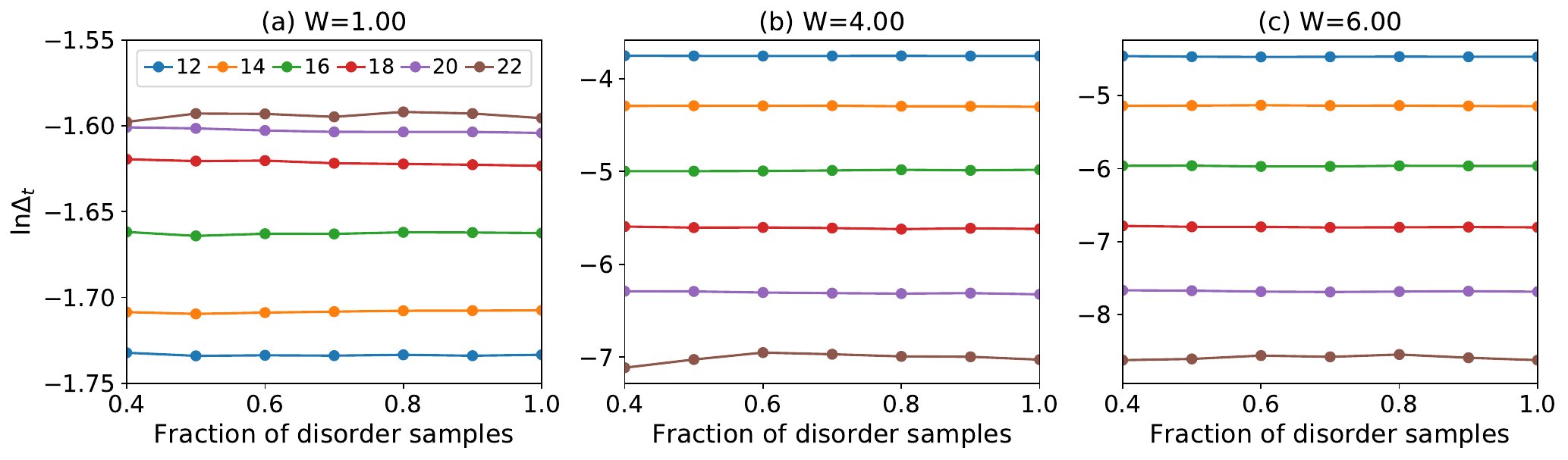}
\caption{Convergence of $\mathrm{ln}\Delta_{t}$ for different number of disorder samples i.e. realizations. 
The maximum number of disorder samples used in these calculations are shown in Table~\ref{Table:disNum}.
%$10000,5000,2000,1000,1050,120$ for system sizes $L=12,14,16,18,20,22$ respectively.
For each system size the number of samples is varied from 0.4 times the maximum sample size to the maximum sample size. The convergence is tested for three disorder values $W=1.0$ (in the thermal phase), $4.0$ (near the transition), and $6.0$ (in the MBL phase). Values of $L$ corresponding to the colors are shown in the legend of (a).}\label{AppFig:convergence}
\end{figure*}
%%%%%%%%%%%%%%%%%%%%%%%%%%%%

\begin{figure*}[h!]
\includegraphics[width=\textwidth]{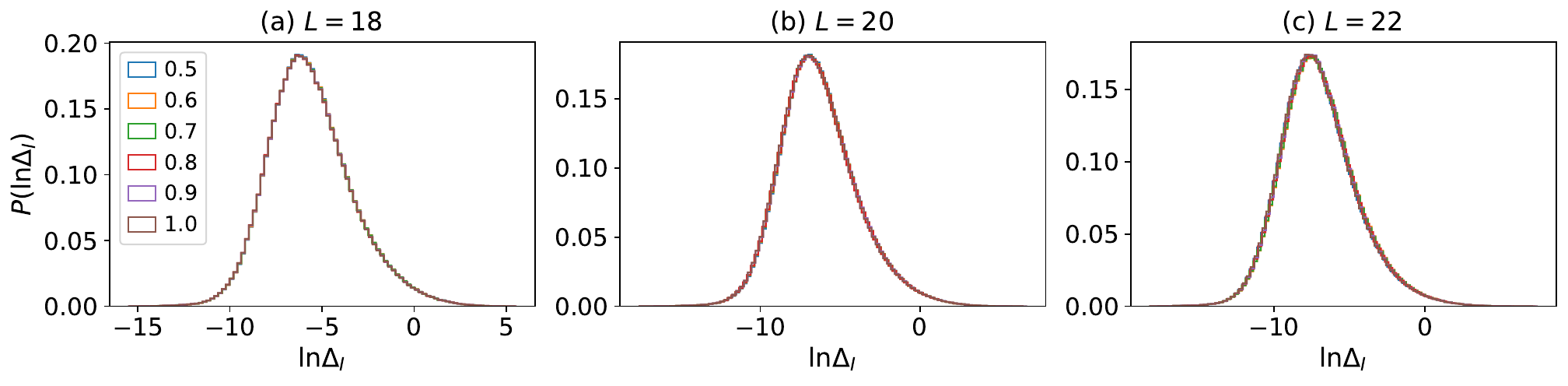}
\caption{
%{\textcolor{red}{Two issues with this figure: (i) the legend cannot be $W$. We need to use a different notation for the fraction of samples such as $f_\mathrm{samp}$. (ii) The caption says this fraction changes from 0.4 to 1 but probably it should be 0.5 to 1.}}
Convergence of the distribution of $\mathrm{ln}\Delta_I$ for different sample sizes at disorder strength $W=4.0$ near the transition. Similar to Fig.~\ref{AppFig:convergence-rel}, the sample size is changed, as shown in the legend of (a), from 0.5 times the maximum sample size to the maximum sample sizes for all system sizes. For all sample sizes and system sizes considered, the distribution is seen to be the same.}
\label{AppFig:distr-convergence}
\end{figure*}

\begin{table}
\begin{tabular}{|c|c|c|}
\hline
$\hspace{0.2cm}L\hspace{0.2cm}$ & \# disorder realizations & \# of total data points \\
\hline
\hline 
12 & $10^4$ & $6\times10^4$ \\
14 & $5\times10^3$ & $8\times10^5$ \\
16 & $2\times10^3$ & $11\times10^5$ \\
18 & $10^3$ & $17\times10^5$ \\
20 & 500 & $27\times10^5$ \\
22 & 130 & $23\times10^5$\\
\hline
\end{tabular}
\caption{\label{Table:disNum} $L$ and corresponding number of disorder realizations employed to generate the distributions of $\di$ at each $W$ point are shown in the first two columns. The third column estimates the product of number of disorder realizations and the number of FS sites in the middle slice.}
\end{table}
%%%%%%%%%%%%%%%%%%%%%%%

However, the quantity $\Delta_I$ is not expected to be self-averaging, i.e., there can be sample-to-sample fluctuations even in the thermodynamic limit. This kind of phenomenon is well known for conductance and its universal sample-to-sample fluctuations in studies~\cite{Lee_PRL1985,Slevin_PRL2001} of single-particle Anderson localization. Still, a scaling theory for the Anderson localization transition can be constructed using the typical or average value of the conductance~\cite{Slevin_PRL2001}. In a similar spirit, we consider the typical value of $\Delta_I$, and find its value to be converged within the disordered realizations considered (Table~\ref{Table:disNum}). 
We have also verified that the distributions do not change further with an increasing number of disorder realizations. 
Fig.~\ref{AppFig:convergence} shows the convergence of $\ln\dt$ for different system sizes at three different disorder strength, $W = 1.0$ (in the thermal phase), 4.0 (near the critical point), and 6.0 (in the MBL phase). The x-axis represents the fraction of the maximum number of disorder samples considered for the calculation of $\ln\dt$ as shown in Table~\ref{Table:disNum}. 
%The maximum accessible disorder samples are 10000, 5000, 2000, 1000, 1050, 120 for system sizes $L =$ 12, 14, 16, 18, 20, 22 respectively. 
For example, for $L = 18$ the disorder sample size is changed from 400 to 1000 in steps of 100. 
%Fig.?? shows that the value of $\ln\dt$ is independent of the sample size. 
%%%%%%%%%%%%%%%%%%%%%%%%%%%%%%

%%%%%%%%%%%%%%%%%%%%%%55
Fig.~\ref{AppFig:distr-convergence} shows the convergence of the distribution of $\ln\di$ over different samples and Fock space sites belonging to the middle slice, for system sizes $L =18$, 20, 22. We show here only the convergence for $W = 4.0$, although the convergence holds for all the disorder strengths. We would also like to emphasize that the typical value $\dt$ for the local FS self-energy $\di$ is obtained by averaging over both disorder realizations and FS sites $I$ in the middle slice of the FS lattice [Fig.~\ref{Fig:Fig1}(a)]. Since the number of sites in the middle slice increases exponentially with system size, the {\it effective} number of samples, albeit not independent, over which statistics is accumulated, is $\sim 10^6$ for $L\ge 14$, see Table~\ref{Table:disNum}. This leads to $\dt$ and $P(\ln\di)$ converging quite rapidly for larger system sizes like $L = 22$ even when the number of independent disorder realizations is smaller for larger $L$, as we demonstrate here.
%%%%%%%%%%%%%%%%%%%%%%%%%%%%

%%%%%%%%%%%%%%%%%%%%%%%%%%%%%
%\medskip
%\bibliographystyle{unsrt}
\bibliography{reference}

\end{document}